# Vaulted Passkeys: A Device-Bound Proposal for Authenticated Credential Export and Import

**Pol Henarejos, Ph.D.**

Centre Tecnològic de Telecomunicacions de Catalunya (CTTC)
Av. Carl Friedrich Gauss 7, 08860, Castelldefels, Barcelona

**Abstract—**Hardware authenticators deliberately resist private-key extraction, yet replacement, disaster recovery, and controlled migration create a legitimate need for portability. Existing guidance for device-bound credentials commonly reduces recovery risk by registering an additional authenticator before failure. That creates an independent credential registration and requires replacement hardware to exist in advance; it is redundancy, not a backup of the original credential. This paper addresses the resulting recovery gap by exporting protected credential state while the source is available and restoring it to hardware acquired later, without cloning a complete authenticator or exposing plaintext private keys to routine desktop software. We propose Vaulted Passkeys, a device-bound architecture in which a random 256-bit Kvault protects authenticated PKV1 credential envelopes through HKDF-separated keys and four explicit AEAD profiles. The design separates enrollment from export/import and the required vault from optional identity. We contribute a role-separated system model, wire format, threat analysis, implementation mapping, and falsifiable evaluation plan. The prototype demonstrates feasibility but is neither a formal security proof nor a proposed final standard.

***Index Terms***—*passkeys, WebAuthn, CTAP2, hardware authenticator, credential migration, authenticated encryption, X448, HKDF, provisioning, key separation, proposal.*

## I. INTRODUCTION

The central promise of a passkey is that authentication does not require a bearer copy of a password or a reusable private key in an application database. A credential private key is generated, stored, and used under the control of an authenticator or credential provider. WebAuthn describes discoverable credentials as credential sources stored on the client side, while CTAP defines authenticator operations for creating, using, and managing such credentials [1, 2]. These specifications intentionally focus on authentication ceremonies and credential management; they do not define a generic command by which a hardware authenticator exports every credential private key to an external application.

That omission is reasonable from a conservative security perspective. A private-key export operation is powerful: an application that can invoke it may turn phishing-resistant credentials into portable key material. Yet the absence of a controlled mechanism creates practical pressure. Organizations need hardware-backed backup and recovery. Users replace or repair tokens. FIDO Alliance recovery guidance recommends registering multiple authenticators to reduce account-recovery needs and, when that is infeasible, repeating identity proofing or onboarding [3]. This is useful redundancy, but it is not a backup of an existing credential: every relying party creates another credential registration, and the spare authenticator must be available and registered before the primary device fails.

The temporal distinction is central. Registration redundancy requires two usable authenticators during the preparation phase—effectively requiring the user or organization to own the spare from day zero, or at least before any loss. Vaulted Passkeys instead permits an authorized export while the source board works, secure offline retention of the PKV1 blob and Kvault recovery material, and acquisition of the replacement board only when it is actually needed. The replacement is then enrolled into the same vault domain and imports the protected credential. The cost is shifted from idle spare hardware to disciplined custody of encrypted recovery artifacts and their passphrase.

Vaulted Passkeys explores a middle position. It does not make export a normal WebAuthn operation, and it does not claim that every authenticator should export keys. Instead, it proposes a gated capability for a specifically enrolled vault. The vault layer is the essential mechanism: Kvault is provisioned into the device, and the device performs PIN-authorized export or import of an opaque, authenticated envelope. The identity layer is separate and optional: when an organization wants to know which authorized provisioning service supplied the vault, a firmware-embedded CA root and board-serial check provide that evidence. The routine operator never receives the plaintext private key; the GUI receives only the protected PKV1 blob and metadata.

### *1.1 Problem Statements and Proposed Responses*

The work is organized around four independently challengeable problems and corresponding design responses:

- Deferred portability—Issue: A replacement board may not exist at export time. Proposed response: PKV1 exports an authorized credential for restoration to a distinct board acquired later, without full-device cloning.
- Role separation—Issue: Routine management should not receive vault roots or plaintext keys. Proposed response: Separate enrollment and mobility ceremonies enforce this boundary.
- Optional identity—Issue: Identity attribution must not become a vault-cryptography dependency. Proposed response: An optional CA layer separately authenticates the provisioning authority and target board.
- Recoverability boundary—Issue: Loss scenarios require defined outcomes. Proposed response: The threat model specifies the consequences of losing or compromising a board, PKV1 blob, Kvault recovery envelope, or passphrase.

### *1.2 Contributions and Claim Boundary*

The paper contributes: (i) a two-layer model separating credential protection from optional provisioning identity; (ii) a two-ceremony model separating the security-critical enroller from the routine credential manager; (iii) the PKV1 envelope and context-separated key schedule; (iv) an implementation across firmware, enroller, GUI, and tests; and (v) a threat model and evaluation protocol intended to make failures observable. The contribution is a design proposal and open implementation artifact. It does not claim universal deployability, cryptographic novelty, formal verification, or measured production readiness.

### *1.3 Reading Guide*

**Plain-language model.** Kvault is the master secret shared by members of one vault domain. PKV1 is a locked package containing one credential. The authenticator PIN/UV token is the authorization gate for creating or opening that package. The optional CA certificate is an identity badge for provisioning; it is not the encryption key and is not required when anonymity is the policy.

**Scope and implementation neutrality.** This paper is a non-normative, company-neutral research proposal. For clarity and reproducibility, it names the PicoKeys open-source ecosystem used by the evaluated prototype: Pico-FIDO is the reference vault-capable authenticator, PicoKeyApp is the reference credential-management client, and pico-vault-enroller is the reference security-critical enroller [4]. These names identify implementations, not required protocol actors or dependencies. Except where implementation-specific behavior is reported, the architecture, roles, wire format, and security arguments may be implemented by independent authenticators, enrollers, clients, and identity authorities. This work is not a FIDO, W3C, IETF, or PicoKeys standard.

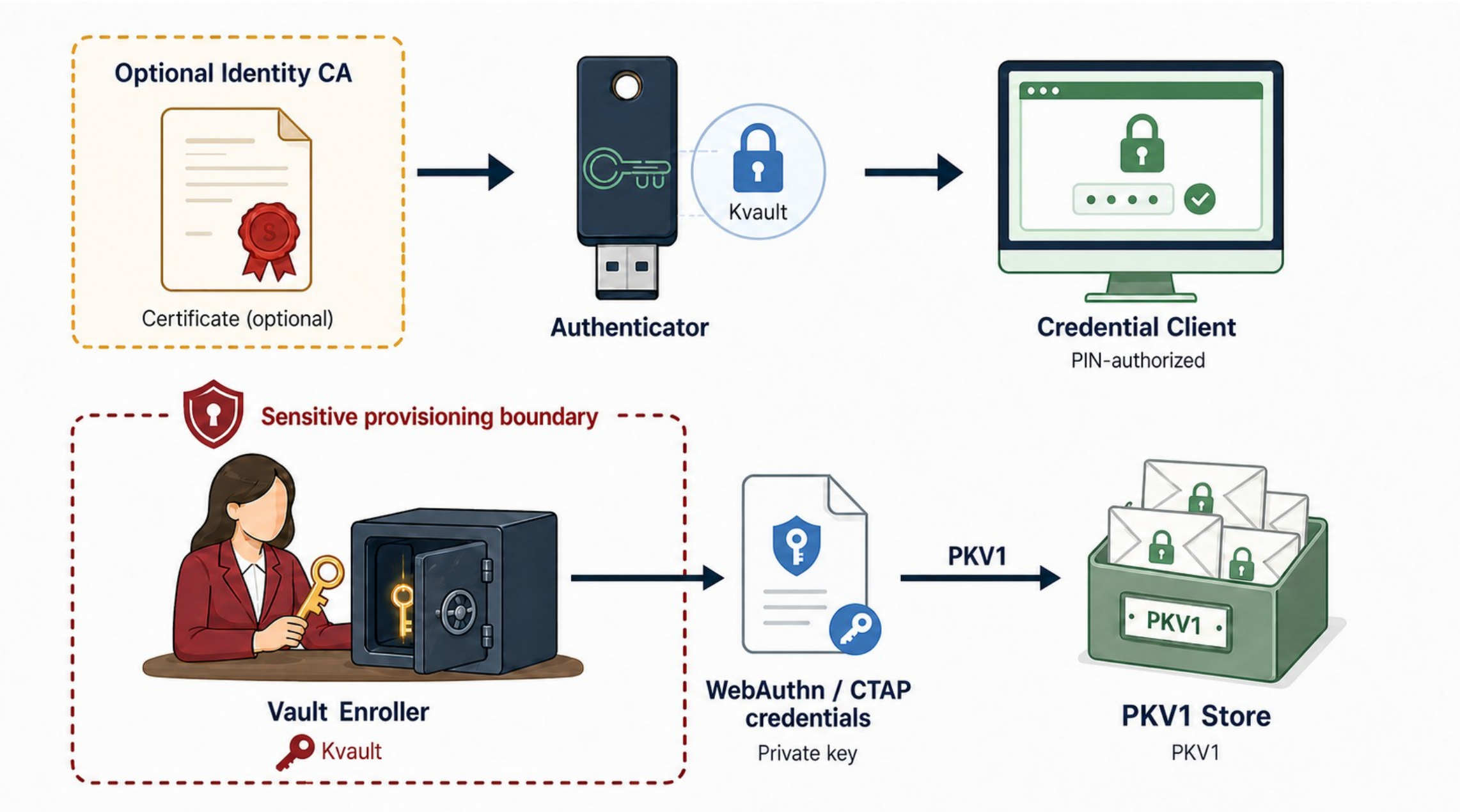

*Figure 1. The proposal separates the optional identity CA, sensitive enroller, hardware authenticator, and GUI credential provider. The red boundary identifies the component that handles plaintext Kvault during provisioning.*

## II. STATE OF THE ART

### *2.1 WebAuthn and CTAP*

WebAuthn scopes a credential to a relying party (RP) and defines browser-mediated creation and assertion ceremonies. A discoverable credential can be selected using an RP identifier without the RP first supplying a credential identifier [1]. CTAP supplies the client-to-authenticator interface and includes credential-management operations for discoverable credentials [2]. Together, these technologies provide a strong authentication substrate, but their ordinary control model is intentionally asymmetric: the RP receives a public key and assertions, while the credential private key remains in the authenticator or provider.

WebAuthn Level 3 nevertheless models backup-capable credentials through the Backup Eligibility (BE) and Backup State (BS) flags and recognizes that backup may occur through mechanisms including manual import/export [1]. Those semantics describe credential properties and relying-party-visible state; WebAuthn explicitly does not define a protocol for backing up private keys or sharing them between authenticators. CTAP 2.3 carries authenticator data and provides credential-management operations to enumerate, inspect, update, and delete discoverable credentials, but it does not transpose the WebAuthn backup model into standardized authenticator commands for exporting or importing a credential private key [2]. This missing client-to-authenticator operation is the specific protocol gap addressed experimentally by PKV1 and the vendor Vault commands.

### *2.2 Credential Exchange and Provider Migration*

The FIDO Alliance Credential Exchange Protocol (CXP) and Credential Exchange Format (CXF) are the closest adjacent work. As of this draft, CXF 1.0 is a Proposed Standard with March 2026 errata, whereas CXP remains a Working Draft. CXP describes provider-to-provider export/import, request and response modes, Diffie-Hellman or HPKE-based protection, and an archive containing credential payloads. CXF provides the credential representation. This proposal shares the goal of avoiding cleartext CSV-style migration and agrees that user or organizational authorization is essential [5, 6].

The difference is the trust boundary. CXP is a provider-interoperability protocol: the exporter and importer are credential providers, and the protocol is designed for online or offline exchange. Vaulted Passkeys is narrower and device-centric. Its exporter is a hardware authenticator command, its root secret is enrolled into that device, and its PKV1 envelope is deliberately bound to a vault identity and board serial. A future implementation could map PKV1 into CXF/CXP, but this paper does not claim that PKV1 is interoperable with CXF.

### *2.3 Cryptographic Building Blocks*

The proposal composes established primitives rather than inventing new cryptography. X448 is used for an enrollment key agreement, HKDF provides context-separated key derivation, and ChaCha20-Poly1305 and AES-GCM provide authenticated encryption with associated data. These choices are grounded in the established specifications for X448 [7], HKDF [8], ChaCha20-Poly1305 [9], and GCM [10]. The research contribution is therefore architectural: how these primitives, certificate identity, device policy, and operational roles are composed for a constrained hardware credential-mobility capability.

### *2.4 Comparative Positioning*

Table I positions the proposal against the two most relevant operational alternatives. The comparison is architectural rather than a claim of superiority in every deployment: provider synchronization may be preferable for consumer convenience, while re-registration is preferable whenever private-key portability is forbidden.

| Property | Full-board clone | Provider exchange / sync | Vaulted Passkeys |
|---|---|---|---|
| **Unit moved** | Complete device state | Provider-defined archive or synchronized credential | One selected credential envelope |
| **Device identity** | May be duplicated | Usually abstracted by provider | Source and target remain distinct |
| **Routine software sees plaintext key** | Clone tooling may see raw storage | Depends on provider boundary | No; GUI handles opaque PKV1 |
| **Relying-party change** | None, but clone semantics are ambiguous | None when ecosystem supports it | None |
| **Recovery blast radius** | Entire board state | Provider account/archive | Selected credential plus its vault domain |
| **Primary dependency** | Hardware/storage equivalence | Provider interoperability and account trust | Enrolled vault, firmware policy, and Kvault recovery |

The literature and standards discussion is intentionally bounded to adjacent specifications and primitive definitions. A systematic review of commercial implementations, patents, hardware backup products, and unpublished vendor mechanisms remains future work; absence from this section must not be read as evidence that no comparable mechanism exists.

## III. MOTIVATION AND DESIGN GOALS

The design is motivated by a tension: portability is useful, but uncontrolled portability weakens the security model that makes passkeys attractive. The proposal treats export as a high-impact administrative capability rather than as a convenience API. Its design goals are the following:

### *3.1 Registration Redundancy Is Not Credential Backup*

FIDO Alliance guidance correctly observes that account recovery can be reduced by registering multiple authenticators [3]. However, this approach creates two independent credential key pairs and two relying-party records. It does not preserve the first credential and cannot reconstruct it. In storage terminology, it is active redundancy: both authenticators must exist, be enrolled with every relevant relying party, and remain governed throughout their lifetimes.

This distinction matters after failure. If the primary board is lost before a second authenticator has been registered, the user cannot derive the missing registrations from the surviving relying-party public keys. The user must invoke each relying party's recovery procedure, repeat identity proofing where required, and create new credentials. Even when a spare exists, adding a new relying party later requires remembering to register both boards again; otherwise protection silently diverges.

A credential backup has a different temporal property: it captures recoverable state at time t0 and permits restoration onto replacement media obtained at time t1. Vaulted Passkeys provides that decoupling. While the source board is healthy, an authorized operator exports selected PKV1 blobs and preserves the Argon2id-protected Kvault recovery envelope [11]. No destination board is required at export time. If replacement later becomes necessary, a newly purchased board is enrolled into the same vault domain and imports the selected blobs.

| Property | Register another authenticator | Vaulted Passkeys backup |
|---|---|---|
| **Credential continuity** | New key pair and RP registration | Preserves the exported credential record |
| **Spare hardware at preparation time** | Required | Not required |
| **Work before failure** | Register the spare separately at every RP | Export selected blobs and protect recovery material |
| **Work after failure** | Use an already-registered spare | Buy a board, enroll it into the vault, then import |
| **Failure if preparation is incomplete** | RP recovery or renewed identity proofing | Unavailable blobs or Kvault material make recovery impossible |
| **Stored recovery asset** | A second live authenticator | Encrypted PKV1 blobs plus protected Kvault envelope and passphrase |

**Core motivation.** Registering a second board protects availability through pre-provisioned redundancy. PKV1 protects availability through deferred restoration. The proposal does not eliminate preparation: it replaces the requirement to own and register spare hardware in advance with the requirement to preserve encrypted credential blobs, the Argon2id-protected Kvault envelope, and its passphrase.

### *3.2 Credential Export Instead of Full-Board Cloning*

A complete board clone is a tempting recovery model: copy all flash, keys, configuration, counters, and credential records to another board. It is also the wrong security boundary for credential mobility. A board is more than a credential container. Its state includes device identity, attestation material, PIN and retry state, firmware configuration, monotonic counters, manufacturing data, registration policy, and potentially unrelated application secrets. Copying that state creates an indistinguishable second device and makes it difficult to reason about identity, revocation, counter continuity, and ownership.

#### *3.2.1 Least-privilege recovery*

PKV1 exports only the credential selected by an authorized user and carries only the metadata required to recreate that credential. The target board keeps its own hardware identity, attestation keys, PIN state, firmware state, and board serial. The resulting recovery operation is therefore narrower: it transfers a credential capability, not an entire security principal. This reduces blast radius when a backup is copied, misplaced, or handled by a different operator.

#### *3.2.2 Identity and lifecycle safety*

Cloning a board can duplicate an identity that relying parties or an organization expect to be unique. It can also roll back counters or replicate state that was intended to change monotonically. Vaulted Passkeys keeps identity and credential mobility separate. The optional identity layer can certify who provisioned a vault and which board was enrolled, but the PKV1 envelope does not clone the board certificate, board serial, attestation identity, or device-internal key-management state.

### 3.2.3 Practical recovery

A full clone is tightly coupled to the exact hardware, firmware layout, storage format, and lifecycle of the source board. A credential envelope is a smaller, inspectable unit with an explicit version, algorithm profile, vault identifier, credential hash, and authenticated metadata. This makes it more suitable for controlled backup, selective restoration, and future format bridging. The trade-off is deliberate: the proposal does not promise a transparent image-level backup of every board function; it promises auditable portability of selected credentials.

### 3.3 Design Goals

- Keep credential private keys out of the clear during normal GUI export/import. The device creates and consumes the protected envelope; the GUI stores and transports opaque bytes.
- Keep the vault layer independent from identity policy. Kvault provisioning and PKV1 mobility are the required mechanism; an optional CA-signed certificate and exact board-serial SAN provide an additional identity check when an organization needs attribution.
- Separate responsibilities. The optional identity CA, sensitive enroller, hardware authenticator, and export/import GUI may be different entities with different operational controls.
- Bind exported data to the intended vault and credential. The vault identifier and credential hash are included in the authenticated envelope header and influence key derivation.
- Support cryptographic agility without silently changing semantics. Four fixed algorithm profiles make the selected protection visible and testable.
- Fail closed at trust boundaries. Invalid certificates, incorrect serials, wrong vault identifiers, malformed envelopes, invalid authentication, and AEAD failures must reject the operation.
- Remain compatible with normal passkey use. Relying parties need not change their WebAuthn ceremonies, and the export capability is not part of the ordinary assertion path.

## IV. SYSTEM MODEL AND ROLES

### 4.1 Assets and Trust Boundaries

The protected assets are credential private keys and metadata, Kvault, the enroller’s recovery state and passphrase, the authenticator’s wrapped vault state, PIN/UV authorization tokens, and—when identity is enabled—the CA signing key and issued certificate. Trust boundaries occur between the optional CA and enroller, between enroller and authenticator during provisioning, between GUI and authenticator during mobility, and between local blob storage and its operator. The design intentionally gives no single routine desktop component every asset.

### 4.2 Roles

The proposal defines roles by security responsibility rather than by product boundary. One organization may operate several roles, but the protocol does not require that they be co-located. This matters because the most sensitive operation—creating or recovering Kvault—should not be casually combined with routine credential export.

The identity CA is optional and is not an application backend: its role is only to sign or reject the certificate used to identify the provisioning authority to the enroller and firmware.

| Role | Primary responsibility | Security significance |
| --- | --- | --- |
| **Optional identity CA** | Signs the certificate for the enroller's X448 public key and board serial, when attribution is required. | Compromise enables unauthorized identity claims; the private CA key must be protected and governed. It is not required for anonymous provisioning. |
| **Enroller (reference: pico-vault-enroller)** | Creates Kvault and the X448 key pair, obtains the optional certificate, performs enrollment, and stores the encrypted enrollment envelope. | Critical provisioning authority; it handles plaintext Kvault and must be open to audit, minimized, and operationally protected. |
| **Vault-capable authenticator (reference: Pico-FIDO)** | Validates applicable identity data, stores Kvault wrapped by a device-internal key, and performs PIN-authorized export/import. | The security boundary that keeps private keys inside the device during routine mobility. |
| **Credential-management client (reference: PicoKeyApp)** | Unlocks credential management, checks applicable board-registration policy, requests export/import, and persists opaque PKV1 blobs locally. | Should not need Kvault or plaintext credential private keys; local storage remains sensitive because it contains recoverable ciphertext. |
| **Relying party** | Uses ordinary WebAuthn public keys and assertions. | Does not participate in this proposal and need not be modified. |

*4.3 Assumptions*

- The authenticator executes the reviewed firmware and its device-internal key storage is not already compromised.
- The platform provides cryptographically suitable randomness for Kvault, X448 keys, challenges, and every AEAD nonce.
- PIN/UV authorization is obtained through the normal CTAP security boundary; an adversary who already controls an authorized session is not prevented from requesting an intentional export.
- The enroller host is trustworthy during the short provisioning ceremony, and operators independently protect its Argon2id envelope and passphrase afterward.
- If the identity layer is enabled, the embedded root, CA issuance process, certificate profile, and board serial source are governed correctly. No such assumption is needed for anonymous vault provisioning.
- Relying parties accept a restored credential because its key pair and credential identifier are preserved; policy-specific RP behavior must still be tested.

*4.4 Non-goals*

- A general API for arbitrary private-key extraction, unattended cloud synchronization, or transparent cloning of flash and device identity.
- Protection against invasive physical extraction, power analysis, electromagnetic analysis, fault injection, malicious firmware replacement, or a compromised build pipeline.
- Recovery when every copy of Kvault or its usable recovery material is lost; deliberate unrecoverability is part of the model.

- A claim that two sequential AEAD layers are automatically stronger than one correctly implemented AEAD, or that PKV1 should replace CXP/CXF.

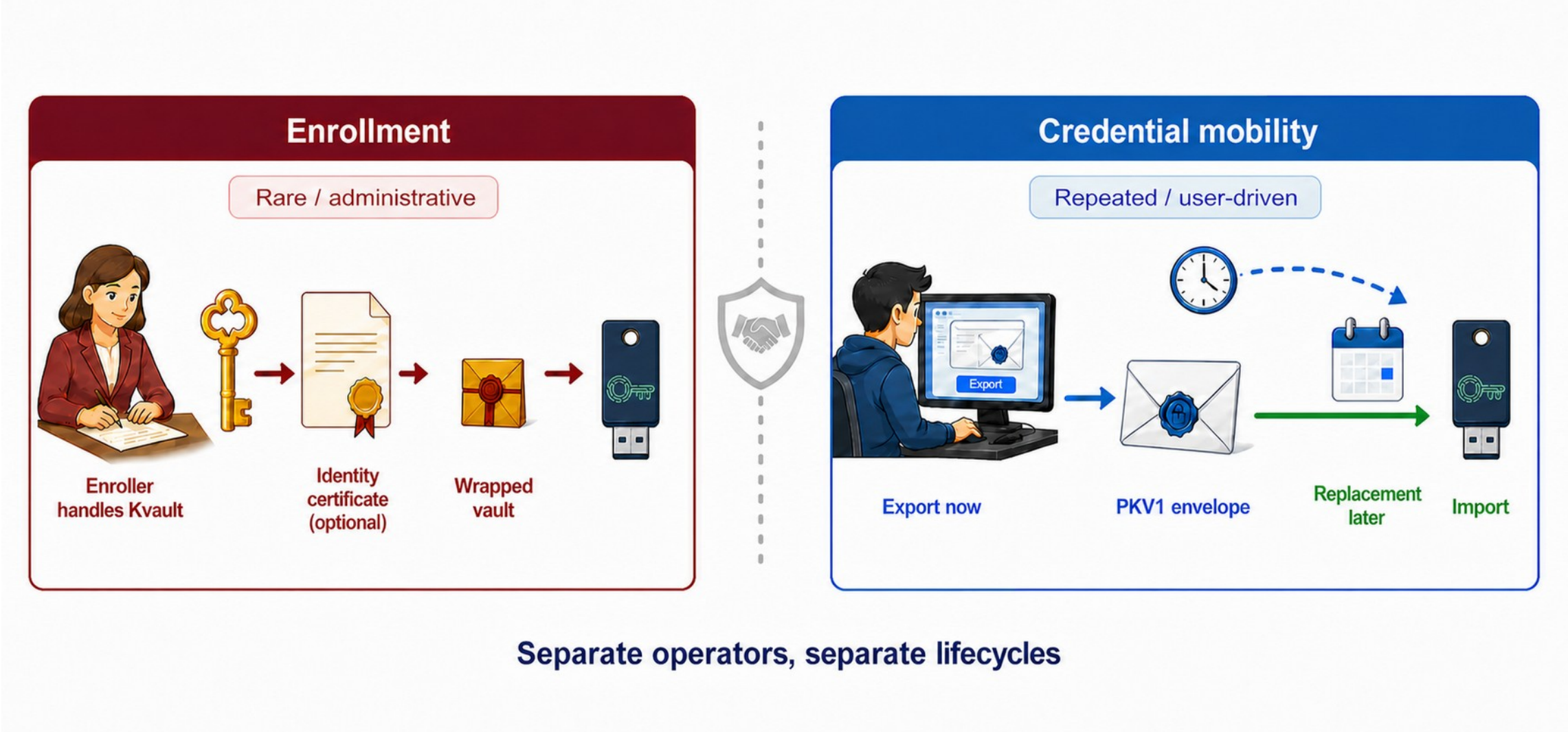

*Figure 2. Enrollment creates durable vault state; export requires no destination board, so replacement hardware can be acquired later.*

## V. PROTOCOL OVERVIEW

The proposal has two layers and two deliberately separate ceremonies. The vault layer is required: it provisions Kvault into the authenticator and uses it to protect selected credential exports. The identity layer is optional: it lets the device and provisioning service check whether a certificate was signed by a trusted organizational CA and whether the certificate names the intended board. Enrollment is a provisioning ceremony. Export/import is a credential-mobility ceremony. The first establishes the device's relationship with a vault; the second uses that established relationship. Treating them as separate processes allows, for example, a security administrator or manufacturing service to enroll a board while a user-facing credential manager later performs export/import under a different policy.

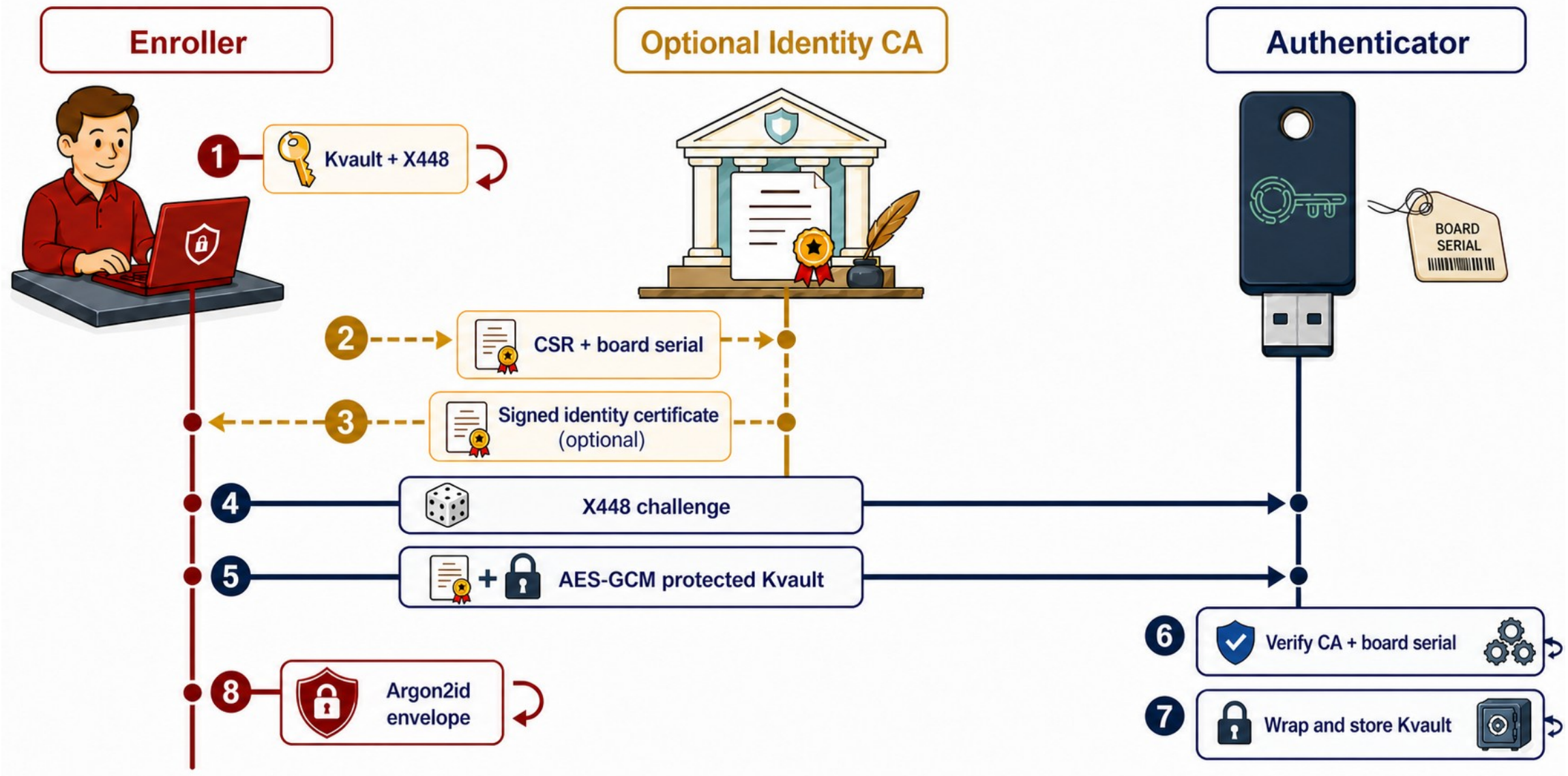

*Figure 3. Enrollment is one-time provisioning: certificate issuance and device acceptance precede vault use.*

### *5.1 The Required Vault Layer*

The vault layer is the cryptographic mechanism that makes credential mobility possible. It does not depend on knowing the human, organization, or service that initiated provisioning. The enroller generates a uniformly random 32-byte Kvault, and the device stores it only after the enrollment packet is authenticated. The resulting vault_id identifies that secret without revealing it. Later, PKV1 derives credential- and layer-specific keys from Kvault and refuses envelopes from another vault.

The vault layer establishes whether a device and an envelope participate in the same protected credential-mobility domain. Attribution of the entity that provisioned the domain is deliberately outside the vault layer and belongs to the optional identity layer below.

### *5.2 The Optional Identity Layer*

The identity layer is an attribution and provisioning-policy mechanism, not a requirement for the vault cryptography. When enabled, the enroller creates an X448 key pair and obtains a certificate signed by an organization's Vault CA. The CA is not a general-purpose backend and does not need to see Kvault, credential contents, PKV1 blobs, or routine export/import traffic. Its narrow purpose is to let the firmware verify that the provisioning certificate was signed by the trusted CA and identifies the intended board.

The firmware validates the certificate chain against a CA root embedded in firmware and checks for an exact board-serial match in the certificate subject alternative name. A certificate signed by another CA, or a certificate whose serial does not match the board, is rejected at enrollment finish. This check identifies the provisioning authority and intended board; it does not itself create the vault, protect PKV1, or authorize later exports. If full anonymity is desired, the identity certificate and CA check may be omitted or replaced by another explicit provisioning policy. The required vault layer remains unchanged.

### *5.3 Enrollment Ceremony*

The enroller generates a uniformly random 32-byte Kvault and an X448 private/public key pair. It computes the 32-byte vault identifier as:

```
vault_id = SHA-256("PicoKeys Vault ID v1" || Kvault)
```

When the optional identity layer is enabled, the enroller sends the X448 public key to the organization's Vault CA. The CA signs a certificate containing the public key in the agreed vault-key extension and the board serial in a DNS-name or URI subject-alternative-name entry. The serial is not merely display metadata: at enrollment finish, the firmware searches the certificate SAN sequence for an exact byte-for-byte match to the board's own serial string. Without this optional layer, the enrollment packet still needs an authenticated provisioning policy, but it need not contain a CA-issued identity certificate.

The device's enrollment-begin command (vendor command 0x05, subcommand 0x02) returns a device ephemeral X448 public key and a fresh 32-byte challenge. The enroller derives an enrollment session key:

```
Z    = X448(enroller_private, device_ephemeral_public)
info = "PicoKeys Vault enrollment v1" || challenge || enroller_public ||
device_ephemeral_public
Ke   = HKDF-SHA256(salt = ∅, IKM = Z, info = info, L = 32)
```

When the identity layer is enabled, the finish packet is certificate-length-prefixed, followed by the DER certificate, a 12-byte AES-GCM nonce, and an AES-GCM ciphertext/tag [10]. The encrypted plaintext is Kvault followed by a one-byte label length and an optional UTF-8 label. The `info` value is authenticated data and also binds the key derivation to the challenge and both public keys. The device performs, in order, packet-length checks, optional X.509 parsing, optional CA-chain and serial-SAN validation, X448 public-key extraction, X448/HKDF derivation, and AES-GCM authentication/decryption. Only after the applicable checks succeed does it commit the vault key to persistent storage. The certificate is validated from the finish packet and is not persisted by the device.

The device stores the vault key encrypted under a device-internal key-management secret. The enroller separately writes an enrollment envelope protected by a user-supplied passphrase. The current implementation derives its 32-byte AES-GCM key with Argon2id using three iterations, four lanes, and 64 MiB of memory, then authenticates the ciphertext with the fixed AAD `PicoKeys Kvault envelope v1` [11]. The envelope contains Kvault, the enroller's X448 private key, the certificate when the identity layer is used, license association, label, and vault identifier. The device therefore does not need the user's vault passphrase, while the enroller can recover the vault state for a later authorized provisioning or recovery operation.

### *5.4 Separation of Enrollment and Export/Import*

This separation is the central operational proposal. Enrollment is sensitive because the enroller sees plaintext Kvault and can create the root of credential-mobility authority. Export/import is sensitive in a different way: it moves recoverable credential material, but it can be performed without exposing Kvault to the GUI. The entities may therefore be split as follows:

1. A provisioning authority operates the open-source pico-vault-enroller and is allowed to handle Kvault only during controlled enrollment.
2. When identity attribution is required, an optional CA operator issues board-scoped identity certificates but does not need to see credential contents or routine export traffic.
3. A user or credential-management operator runs PicoKeyApp and receives only authenticated PKV1 blobs from the device.
4. The relying parties remain unchanged and continue to see normal WebAuthn registration and assertion behavior.

**Why the enroller must be open source.** The enroller is the component that receives or generates plaintext Kvault, derives the Argon2id passphrase-protected envelope, handles the optional CA certificate, and initiates device enrollment. A closed or opaque enroller would create an unreviewable plaintext-key sink. The proposed deployment therefore treats the separate pico-vault-enroller project as a security-critical, auditable component: its source, dependencies, build process, and release artifacts should be reviewable, reproducible where practical, and kept smaller than the GUI application.

## VI. PKV1 CREDENTIAL ENVELOPE

After enrollment, the device exposes two new high-impact operations: export (0x04) and import (0x05). Both require a valid CTAP PIN/UV authorization with authenticator-configuration permission. The export request identifies a credential and selects one of four algorithm profiles. The device locates the resident credential, reconstructs or loads its private key, gathers metadata, and encrypts a CBOR payload.

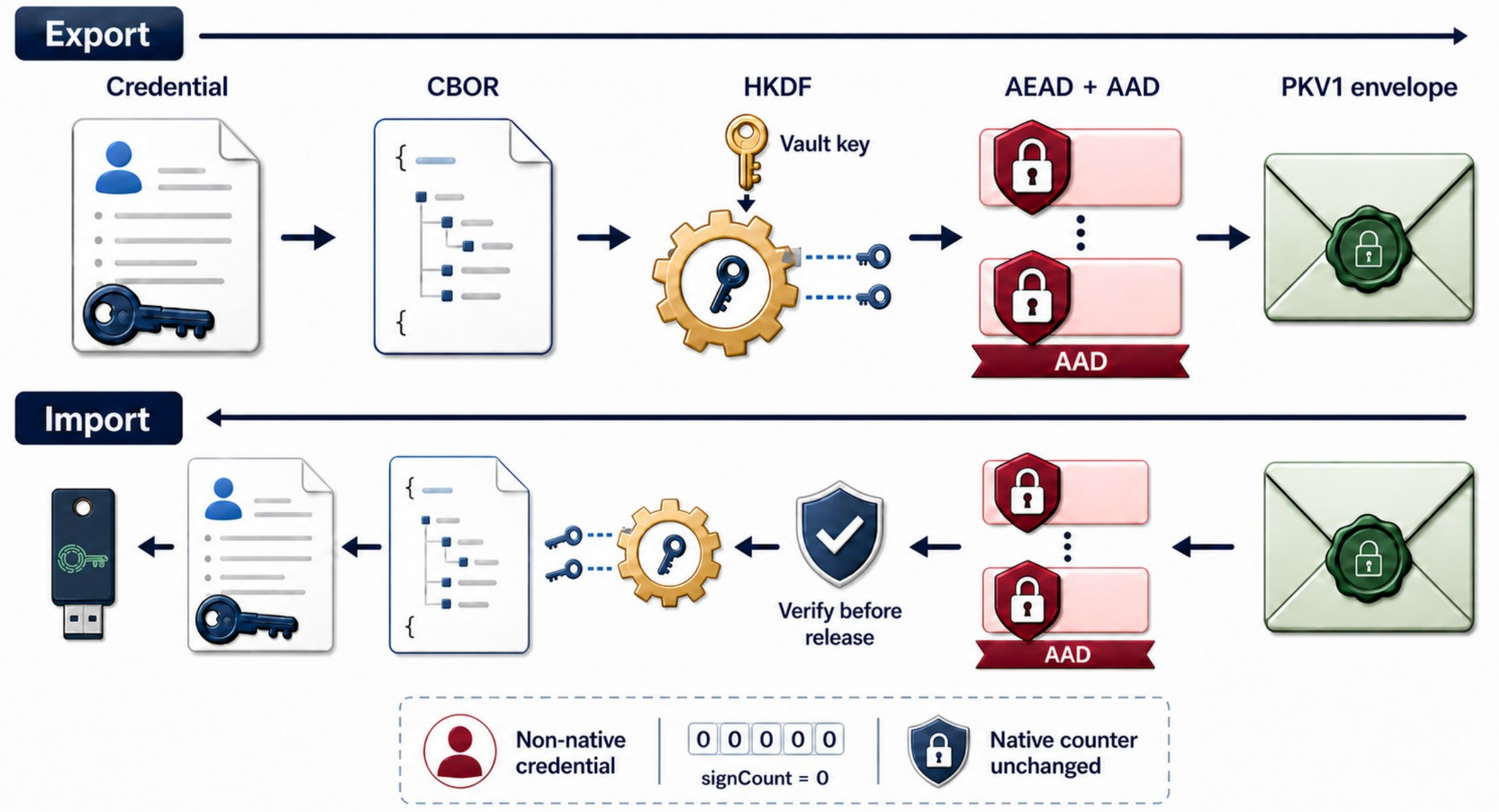

*Figure 4. PKV1 is an authenticated, self-describing envelope. Import reverses the layers and validates before credential recreation.*

### *6.1 Header and Payload*

The current PKV1 header is 86 bytes. Multi-byte quantities are not needed in the fixed fields shown below; the serial is length-prefixed and padded to a 16-byte maximum. The complete header is passed as AEAD associated data, so changing an algorithm identifier, vault identifier, credential hash, board serial, or magic/version value invalidates decryption.

```
offset  size  field
0        4    magic = 50 4B 56 01 ("PKV1")
4       32    vault_id = SHA-256(domain || Kvault)
36      32    credential_hash = SHA-256(requested credential ID)
68       1    serial_len
69      16    board serial bytes, zero-padded
85       1    algorithm ID (1..4)
86       n    12-byte nonce per AEAD layer
86+n     *     ciphertext followed by one 16-byte tag per layer
```

The body plaintext is a six-entry CBOR map. Integer labels keep the encoding compact and unambiguous:

| Key | CBOR type | Field | Semantics |
|---|---|---|---|
| **1** | unsigned integer | version | Credential-plaintext format version; currently 1. |
| **2** | byte string | credential_id | Stored credential identifier recreated on import. |
| **3** | byte string | private_key | Authenticator-private serialization of the credential key material; it remains inside the authenticated ciphertext. |
| **4** | text string | rp_id | Relying-party identifier associated with the credential. |
| **5** | byte string | metadata | A second CBOR map, encoded as bytes, containing the fields needed by credential import. |
| **6** | byte string | requested_id | Exact export input; its SHA-256 digest is the header credential_hash and enters layer-key derivation. |

The distinction between keys 2 and 6 is intentional. Export lookup may accept either the stored credential identifier or another identifier recognized by the authenticator's resident-credential matching logic. Key 2 preserves the canonical stored identifier; key 6 records the exact input that selected it and cryptographically binds that selection to the header.

The byte string under key 5 decodes to the following metadata map. The map may use indefinite-length CBOR encoding; omitted values retain the stated defaults.

| Key | CBOR type | Field | Presence and meaning |
|---|---|---|---|
| **1** | text string | rp_id | Optional RP identifier. |
| **2** | byte string (32 bytes) | rp_id_hash | Required SHA-256 RP identifier hash used to index and recreate the credential. |
| **3** | byte string | user_id | Optional WebAuthn user handle. |
| **4** | text string | user_name | Optional user name. |
| **5** | text string | user_display_name | Optional display name. |
| **6** | unsigned integer | board_creation | Required implementation creation value retained with the credential. |

| Key | CBOR type | Field | Presence and meaning |
|---|---|---|---|
| **7** | map | extensions | Optional extension state: credBlob, credProtect, hmac-secret, largeBlobKey, and thirdPartyPayment. |
| **8** | boolean | use_sign_count | Required source-credential preference. For a non-native imported credential, imported status overrides this value: assertions report signCount = 0 and do not advance the destination board's native counter. |
| **9** | integer | alg | Optional COSE algorithm identifier [12]; omission denotes ES256. |
| **10** | integer | curve | Optional curve identifier; omission denotes P-256. |
| **11** | map | options | Optional credential options; currently rk as a boolean. |
| **12** | unsigned integer | rtc_creation | Optional real-time-clock creation value. |
| **13** | byte string | resident_id | Optional stable resident identifier used for deterministic key reconstruction. |

Extension key 7 can carry credBlob as a byte string, credProtect as an unsigned integer, hmac-secret as a boolean, and true-valued largeBlobKey or thirdPartyPayment flags. The importer requires all six outer fields and metadata keys 1 (rp_id) and 2 (rp_id_hash). It rejects duplicate known fields, unsupported plaintext versions, invalid bounds, a requested_id whose SHA-256 digest differs from the authenticated header credential_hash, an outer RP ID that differs from metadata, an RP hash that differs from SHA-256(rp_id), and inconsistent algorithm, curve, or private-key material. These checks complete before slot selection or persistent writes. The reference credential-management client, PicoKeyApp, transports PKV1 as opaque bytes and does not receive this cleartext record or Kvault. Separately returned presentation metadata must not be confused with the encrypted metadata copy inside PKV1.

### *6.2 Key Derivation and Algorithm Profiles*

For each credential and layer, the device derives a fresh 32-byte AEAD key from Kvault. The derivation context includes the credential hash, selected algorithm, and layer index:

```
K(layer) = HKDF-SHA256(
    salt = vault_id,
    IKM  = Kvault,
    info = "PicoKeys Vault enrollment v1" ||
           credential_hash ||
           algorithm || layer,
    L    = 32)
```

| ID | Profile | Construction |
|---|---|---|
| **1** | ChaChaPoly | One ChaCha20-Poly1305 layer [9]; 96-bit nonce; 128-bit tag. |
| **2** | AES-GCM | One AES-256-GCM layer [10]; 96-bit nonce; 128-bit tag. |
| **3** | ChaChaPoly + AES-GCM | Plaintext → ChaCha20-Poly1305 → AES-256-GCM, with independent layer keys and nonces. |
| **4** | AES-GCM + ChaChaPoly | Plaintext → AES-256-GCM → ChaCha20-Poly1305, with independent layer keys and nonces. |

The two-layer profiles are offered as an explicit proposal choice, not as a claim that sequential AEAD automatically doubles security. Their benefits are operational: they provide algorithm agility, allow deployments to combine a software-friendly primitive with a hardware-accelerated primitive, and make the envelope exercise independent integrity checks. Their composition must nevertheless be reviewed as a whole, including nonce generation, failure handling, side channels, and downgrade policy.

### *6.3 Reference Processing Algorithms*

The following pseudocode states the security-relevant order of operations. It is descriptive: concrete CBOR keys, bounds, status codes, and zeroization remain implementation requirements.

```
EXPORT(credential_id, profile, pin_uv):
    authorize(pin_uv); require enrolled_vault and valid_profile
    record = load_resident_credential(credential_id)
    header = PKV1(vault_id, SHA256(credential_id), board_serial, profile)
    plaintext = CBOR(record.metadata, record.private_key, credential_id)
    for layer in profile_order(profile):
        key = HKDF(Kvault, vault_id, credential_hash || profile || layer)
        plaintext = AEAD_ENCRYPT(key, fresh_nonce(), header, plaintext)
    erase(record.private_key, plaintext_working_buffers, derived_keys)
    return header || nonces || ciphertext_and_tags
```

```
IMPORT(envelope, pin_uv):
    authorize(pin_uv); parse_bounded_header(envelope)
    require enrolled_vault and header.vault_id == local_vault_id
    ciphertext = envelope.body
    for layer in reverse_profile_order(header.profile):
        key = HKDF(Kvault, vault_id, credential_hash || profile || layer)
        ciphertext = AEAD_DECRYPT(key, nonce[layer], header, ciphertext)
    record = parse_and_validate_authenticated_CBOR(ciphertext)
    commit_credential_atomically(record)
    erase(record.private_key, plaintext_working_buffers, derived_keys)
```

### *6.4 Import Validation and Commit*

1. Require a non-empty envelope, correct PKV1 magic, a valid algorithm identifier, and sufficient bytes for the declared layer count.
2. Load the device's Kvault and recompute vault_id. Reject if the envelope vault identifier does not match the enrolled device vault.

3. Derive the same per-credential, per-layer keys and decrypt in reverse layer order using the complete header as associated data.
4. Parse the authenticated CBOR object once and require all six outer fields plus metadata RP ID and RP hash.
5. Reject duplicate known fields; validate version, bounds, requested-ID/header hash, outer/metadata RP ID, RP ID/hash, and algorithm/curve/private-key consistency; derive the public key and client identifier; then publish all final resident objects through one authenticated multi-object container update.

**Binding nuance.** The PKV1 serial is authenticated and preserved for provenance and GUI policy. The current device import path primarily enforces vault_id equality; a deployment that requires same-board-only imports should add an explicit serial comparison at import. The proposal keeps these policies distinct: vault equality is the cryptographic portability boundary, while board serial equality is an optional stricter administrative boundary.

### *6.5 Imported-Credential Counter Semantics*

A device-wide signature counter is authenticator state, not portable credential key material. Copying its value into a replacement board would overwrite unrelated native state or falsely claim monotonic continuity between independently operating authenticators. PKV1 therefore does not transplant the source board's global counter. Once imported, the credential is marked non-native: every assertion for that credential conveys signCount = 0, and producing that assertion does not read, reset, or increment the destination board's native global counter. Native credentials on the destination continue to use that counter normally.

This rule avoids manufacturing counter continuity that the protocol cannot guarantee, but it cannot erase state already retained by a relying party. If the relying party previously observed a non-zero counter from the source authenticator, a later zero may be treated as a counter anomaly or clone signal. Consequently, signCount = 0 states that counter-based clone detection is unavailable for the imported credential; it does not guarantee seamless acceptance. The current firmware enforces the rule in its assertion path by selecting zero for imported credentials and skipping persistent counter advancement.

## VII. SECURITY ANALYSIS

### *7.1 Adversary Model*

We consider an adversary who can copy, replay, truncate, reorder, or modify files and protocol messages; obtain a lost PKV1 file; present a certificate from an untrusted CA or for another board; operate a normal GUI without Kvault; steal a powered-off board; or attempt malformed-input attacks against parsers. We also consider a malicious or compromised enroller because it observes plaintext Kvault. The model does not assume that open source prevents compromise; it enables inspection and independent builds. Adversaries with valid PIN/UV authorization, arbitrary firmware execution, invasive physical access, CA signing-key control, or control of the enroller during provisioning cross explicit trust assumptions and are treated as residual-risk cases rather than cryptographically excluded attackers.

### *7.2 Security Properties and Claim Status*

The properties below are design objectives supported by construction and implementation tests, not theorem-backed guarantees. Each should eventually be paired with test vectors, negative cases, code review, and—where appropriate—formal analysis.

- Confidentiality of credential private keys against a passive transport observer, a stolen PKV1 blob, and a GUI that only handles opaque export data.
- Integrity and authenticity of the vault envelope, including its algorithm, vault, credential, serial, and ciphertext fields.
- When the identity layer is enabled, only a certificate chaining to the embedded Vault CA and containing the exact board serial is accepted at the finish ceremony.
- Authorization of export/import: operations require an authenticated PIN/UV token with the required permission.
- Key separation: the vault root is not reused directly as an AEAD key; credential and layer context are included in HKDF info.
- Operational separation: routine export/import does not require the GUI to possess Kvault or the enroller passphrase.

### *7.3 Threats and Residual Risk*

| Threat | Existing mitigation | Residual risk / required policy |
|---|---|---|
| **Modified or replayed PKV1 blob** | AEAD tag, full-header AAD, vault_id, credential hash, and structural parsing. | Do not disable authentication errors or import before successful parsing. |
| **Wrong board or rogue certificate** | When enabled, firmware-embedded CA-root verification plus exact board-serial SAN at enrollment finish. | The identity check is optional; if enabled, CA private-key compromise and certificate lifecycle remain critical operational risks. |
| **Stolen GUI vault file** | Private key is encrypted in PKV1; local file is written atomically with restrictive permissions. | A valid enrolled device and PIN can still recover the credential; protect the device and PIN. |
| **Compromised enroller** | Open-source, separate, auditable project; passphrase-protected enrollment envelope. | A malicious enroller sees plaintext Kvault and can provision or recover the vault; keep it isolated and reviewable. |
| **Unauthorized export request** | PIN/UV authorization and GUI board-registration policy. | A user with valid authorization can intentionally export; organizational policy must define who may do so. |
| **Credential leakage in firmware memory** | Zeroization of Kvault, derived keys, plaintext buffers, and private-key buffers on normal paths. | Side-channel, crash-dump, debug, and fault-injection resistance require independent assessment. |
| **Counter discontinuity after import** | Imported non-native credentials report signCount = 0 and do not change the destination's native counter. | A relying party that stored a non-zero source count may flag the transition; PKV1 cannot rewrite relying-party state or preserve global-counter continuity. |

### *7.4 Asset-Loss and Theft Scenarios*

The threat model must distinguish the board, the credential blob, and Kvault. They are different assets with different recovery consequences. Treating them as interchangeable is one of the main reasons a full-board clone is unsafe.

| Scenario | What the cryptography does | Operational consequence |
|---|---|---|
| **Board stolen** | The thief obtains the device and any resident credentials, but not the plaintext Kvault stored by the device. Credential use or export still requires the device's authorization policy, including PIN/UV where configured. | Treat the board as compromised: revoke or replace its relying-party credentials according to organizational policy, protect the PIN, and do not assume the optional identity certificate is a revocation mechanism. A stolen board must not be treated as a harmless lost USB stick. |
| **Board broken or unavailable** | The board cannot perform another export. A replacement can be provisioned only if the enroller's protected enrollment state is recoverable and the deployment issues whatever new identity material its policy requires. | The current prototype does not provide automated disaster recovery. Keep a tested, offline recovery copy of the enroller envelope and its passphrase, or be prepared to re-register credentials with relying parties. |
| **PKV1 credential blob lost** | The blob is not the only copy while the source board remains usable: an authorized operator can export the selected credential again. A lost blob alone does not reveal the private key. | If the source board is also unavailable, the credential cannot be recovered from that blob. Use authenticated, access-controlled backups for blobs and record which board/vault policy applies. |
| **Kvault lost** | PKV1 decryption fails because vault_id and the per-credential keys cannot be reconstructed. The device's wrapped copy and the enroller's Argon2id-protected envelope are the intended independent storage locations. | If both copies are lost—or the enroller envelope cannot be opened because its passphrase is lost—PKV1 blobs are intentionally unrecoverable. Relying parties must then receive newly registered credentials. |

**Recovery invariant.** Kvault is not a password-derived key inside PKV1. It is a high-value 256-bit vault root. The enroller protects its recovery copy with a passphrase-derived Argon2id key and AES-GCM; the board protects its internal copy with device key management. Losing the passphrase-protected recovery envelope, its passphrase, and the enrolled board together is a deliberate cryptographic loss condition, not a recoverable import error.

### *7.5 Security Argument*

The proposal does not argue that export is harmless. It argues that a capability that may exist in practice should be made explicit, narrow, auditable, and cryptographically bounded. When enabled, a board-scoped CA certificate prevents an identity claim signed by an untrusted authority from being accepted by an unrelated board. A vault identifier prevents a blob from one vault being imported into another. Credential-specific HKDF contexts prevent one credential's key from being reused for another. AEAD associated data prevents silent rewriting of identity and policy fields. PIN authorization forces the user or policy engine to cross a visible authorization boundary. Finally, role separation limits how many components can ever observe the root vault secret.

## VIII. IMPLEMENTATION MAPPING

The prototype is implemented across three cooperating codebases. The firmware contains the cryptographic and credential operations. The standalone enroller performs provisioning. The GUI presents the user-facing Vault panel and stores opaque envelopes. The following mapping is intended to make the proposal reproducible and reviewable.

| Component | Relevant operation | Implementation responsibility |
|---|---|---|
| **pico-fido firmware** | Vendor vault subcommands 0x01..0x06 | Certificate parsing and verification, serial SAN check, X448/HKDF enrollment key agreement, persistent wrapped Kvault, PKV1 export/import, imported-counter isolation, and PIN authorization. |
| **pico-vault-enroller** | 0x02 begin / 0x03 finish | Argon2id passphrase-protected envelope, Kvault/X448 generation, optional CA certificate retrieval, packet construction, enrollment state and label handling. This is the critical provisioning entity. |
| **PicoKeyApp GUI** | 0x04 export / 0x05 import | Passkeys unlock state, exact board-registration gate, algorithm selector, local 0600 VaultStore, opaque blob grouping, import/export feedback. |
| **Optional identity CA** | Certificate issuance | Signs the X448 identity certificate and embeds the board serial in the SAN; does not need to participate in routine credential mobility. |

The implementation uses the board registration license as a GUI policy gate: the board serial must be present in the license's board list before the Vault panel reports an enrolled, usable vault. This is intentionally a GUI authorization layer and should not be confused with the firmware's certificate validation. Firmware enrollment establishes cryptographic trust; PicoKeyApp registration establishes whether that board is permitted to expose the feature in the product workflow.

## IX. RESEARCH METHOD AND EVALUATION

### *9.1 Method*

This work follows an artifact-centered security-design method: define the trust boundaries and desired properties, construct the smallest interoperable prototype, map each property to observable tests, and retain negative evidence and limitations. The evaluation unit is not only the ciphertext format; it is the end-to-end ceremony across enroller, firmware, GUI, persistent storage, and recovery. Reproducibility therefore requires source revisions and environment details for all cooperating repositories.

### *9.2 Evaluation Criteria and Metrics*

| Criterion | Observable method | Acceptance criterion |
| --- | --- | --- |
| **Correctness** | Export, delete, import, enumerate, and authenticate with profiles 1–4. | Credential identifier and behavior survive every round trip. |
| **Tamper resistance** | Mutate each authenticated header field, nonce, ciphertext, tag, CBOR length, and algorithm ID. | Every mutation is rejected before persistent credential commit. |
| **Separation** | Instrument/log process boundaries and inspect stored artifacts. | GUI receives no plaintext private key or Kvault; CA receives neither Kvault nor PKV1. |
| **Failure atomicity** | Interrupt or inject failures at parse, decrypt, and commit boundaries. | No partial credential and no reusable plaintext buffer remain. |
| **Cost** | Measure envelope bytes, export/import latency, peak RAM, and code-size delta on emulator and hardware. | Report distributions and hardware/toolchain context; no threshold is claimed yet. |
| **Recovery** | Export without a destination, then provision a fresh board; also exercise blob loss, wrong vault, wrong passphrase, and restored enroller state. | Deferred restore succeeds only with the correct blob and Kvault recovery material; all loss cases match the threat model. |
| Counter isolation | Record the destination native counter, assert repeatedly with an imported credential, then assert with a native credential. | Imported assertions always return zero; they neither reset nor increment the destination native counter, which remains usable by native credentials. |

### *9.3 Current Evidence*

A standardization-oriented proposal should distinguish implemented checks, executable tests, and measurements not yet collected. The current prototype evidence is organized as follows:

- Deterministic envelope tests: derive vault identifiers, construct PKV1 envelopes for all four profiles, decrypt them, and verify credential and serial fields.
- Negative tests: tamper with enrollment ciphertext, PKV1 headers, ciphertext, algorithms, serial lengths, and vault keys; each case must fail authentication or structural validation.
- Firmware build evidence: compile the emulation target with the Vault implementation included; the current development branch builds successfully.
- Live emulator/hardware test: create a real credential, export it under algorithm IDs 1–4, delete it, import each corresponding blob, enumerate the credential, and verify the credential identifier. The test is now written as a four-iteration round trip; it requires an enrolled emulator or physical device and therefore cannot be treated as complete merely because the test code exists.
- Counter isolation: firmware inspection confirms that imported credentials are persistently marked; the assertion path writes a zero counter and skips the persistent counter update. A live test should additionally verify repeated imported assertions and the unaffected native-counter value.
- Import atomicity: the resident-container fault-injection test interrupts every imported-creation persistence event and verifies after reboot that the credential is either absent or complete with credential ID, RP hash, public key, private key, metadata, and imported state.

- GUI evidence: verify registration gating, unlock-state rendering, opaque local persistence, algorithm display, duplicate handling, and atomic multi-credential writes.
- Independent review: audit the enroller first, then review firmware parsing, memory zeroization, error paths, and certificate lifecycle policy.

**Current prototype evidence.** At the time of this draft, the emulation firmware build passes; 21 non-live Vault tests pass; the GUI VaultStore tests pass; and the live all-algorithm test is present but requires an enrolled emulator or hardware board for execution. This distinction is important: a test harness that skips because no vault is enrolled is evidence of correct precondition handling, not evidence of a successful end-to-end migration.

*9.4 Reproduction Protocol*

1. Record the exact commit of pico-fido, pico-vault-enroller, pico-keys-sdk, and picokeyapp, including submodule state and local patches.
2. Build the emulator and hardware target with Vault enabled; record compiler, SDK, crypto-library, board, and configuration versions.
3. Provision a non-production vault with the standalone enroller. If testing identity, use a test CA and certificate containing the exact test-board serial; never publish production CA material.
4. Run deterministic and negative pytest selections first, then the live-marked four-profile round trip on an enrolled emulator and physical board.
5. Publish machine-readable test logs, skip reasons, timings, envelope sizes, and a sanitized failure corpus. Report skipped live tests separately from passes.

## X. LIMITATIONS AND OPEN ISSUES

- PKV1 is not a standard and is not currently a drop-in CXP/CXF archive. Interoperability with another credential provider would require a format bridge or future alignment.
- WebAuthn Level 3 defines backup eligibility and backup state and acknowledges manual import/export as a possible backup mechanism, but it does not define the private-key backup or sharing protocol. CTAP 2.3 does not transpose those semantics into standard authenticator export/import commands: credential management can enumerate, inspect, update, and delete credentials, but not portably export or recreate their private keys [1, 2]. PKV1 therefore remains a vendor-proposal extension rather than a standards-compatible CTAP backup implementation.
- Deferred hardware purchase is possible only if the credential was exported while the source board was operational and both the PKV1 blob and usable Kvault recovery material remain available. The proposal cannot retroactively recover an unexported credential from a lost or destroyed board.
- The current local GUI store is a JSON index containing opaque PKV1 blobs; it is not yet a user-facing portable archive format with explicit version negotiation, multi-vault policy, or cross-platform transfer UX.

- Reporting signCount = 0 for imported credentials prevents false transplantation of a device-wide counter, but a relying party that already stored a non-zero count may still report a suspected clone or require account recovery. The protocol cannot update relying-party state.
- A stolen PKV1 blob is not sufficient to decrypt the credential, but a device with the corresponding enrolled vault and an authorized PIN can import it. Organizational policy must therefore define export authorization, device custody, PIN management, and recovery.
- When the optional identity layer is enabled, its CA root is embedded in firmware. Root rotation, certificate revocation, expiry policy, manufacturing resets, and field replacement procedures require a lifecycle design beyond the current prototype.
- The serial is an identity and provenance binding, not a complete anti-cloning mechanism. A serial-only policy depends on the authenticity of board identity and on protection of the certificate issuance process.
- The dual-layer profiles require careful review of nonce generation, failure atomicity, timing behavior, and the security meaning of composition. They should not be presented as automatically stronger than a single well-deployed AEAD.
- The design has not yet received a formal cryptographic proof, side-channel evaluation, fault-injection assessment, fuzzing campaign, or independent code audit. These are prerequisites for a high-assurance claim.
- The enroller is intentionally the most sensitive software component. Open source improves auditability but does not by itself prevent a compromised build host, malicious dependency, injected runtime, or operator mistake.

## XI. PROPOSED DIRECTIONS FOR A STANDARDIZATION FORUM

The proposal is suitable for discussion if its claims remain narrow. We propose the following concrete directions for a forum or working group:

- Capability boundary: Standardize separately authorized authenticator mobility as a capability distinct from provider-level migration.
- Identity binding: Require profiles to authenticate a defined board serial, attestation identity, manufacturer certificate, organization-issued identifier, or combination. The prototype uses an exact board-serial binding.
- Provenance and import policy: Authenticate provenance and explicitly select same-device, same-organization, or same-vault enforcement. PKV1 authenticates the source serial and primarily enforces vault equality.
- CXF alignment: Define a device-centric CXF mapping that does not require a general-purpose private-key export API.
- Enrollment-authority assurance: Require auditable source, reproducible-build evidence, dependency controls, and a documented key ceremony for authorities handling a vault root.
- Imported-counter semantics: Standardize the relying-party-visible counter transition. This proposal reports signCount = 0 and suppresses increments for imported non-native credentials.

- Algorithm agility: Require explicit, downgrade-resistant selection through fixed approved suites or a standardized HPKE/CXP mechanism.

**Position.** The strongest standardization argument is not that PKV1 is the final format. It is that enrollment, device identity, authorization, key separation, and provider responsibilities should be specified together. A format without a trust model makes export easy to misuse; a trust model without a usable envelope leaves operators with unsafe workarounds.

## XII. CONCLUSIONS

Vaulted Passkeys proposes a practical security architecture for a difficult capability: moving hardware-backed passkey credentials without turning routine GUI operation into plaintext private-key handling. Its motivating distinction is that registering a second authenticator is availability through pre-provisioned redundancy, not a backup of an existing credential. PKV1 instead enables deferred restoration: export while the source is healthy, preserve encrypted recovery material, and acquire the replacement board later. Its architectural idea is separation. A sensitive enroller creates and provisions the vault; when identity attribution is required, an optional CA-signed certificate binds that provisioning claim to a board. The firmware is the cryptographic execution boundary, while the GUI stores authenticated opaque envelopes.

The proposal is intentionally conservative about what it claims. It uses established cryptographic primitives, leaves ordinary WebAuthn relying parties unchanged, and does not present PKV1 as an internet-wide standard. Its novelty is the composition of device-held vault state, optional certificate identity, context-separated AEAD layers, and explicit operational separation. The approach is promising for organization-managed hardware authenticators and controlled recovery, but its security depends on the enroller, optional identity CA when used, firmware, PIN policy, and lifecycle controls.

The next step is not to declare the format finished. It is to make the design reviewable: publish the enroller source and build process, run the full four-profile live round trip on enrolled hardware, add formal negative and fuzz testing, define certificate lifecycle and same-device policy, and examine whether a future bridge to FIDO credential exchange formats is desirable. With those caveats made explicit, the proposal offers a credible basis for discussion in a standardization forum: secure portability as an opt-in, device-bound, auditable capability rather than an invisible weakening of passkey storage.

## APPENDIX A. COMMAND AND DATA SUMMARY

The table below summarizes the prototype vendor-vault command surface. Command numbers are implementation identifiers for the proposal and should not be confused with an assigned standard code point.

| Subcommand | Name | Purpose |
|---|---|---|
| 0x01 | Vault status | Returns vault identifier, enrollment status, button state, device time, and label. |
| 0x02 | Enrollment begin | Generates device ephemeral X448 public key and challenge. |
| 0x03 | Enrollment finish | Validates the inline certificate identity and serial, decrypts the enrollment packet, and stores Kvault; the certificate is not persisted. |
| 0x04 | Export opaque | Exports a selected resident credential as a PKV1 envelope under algorithm ID 1–4. |
| 0x05 | Import opaque | Authenticates, decrypts, parses, and recreates a credential from PKV1. |
| 0x06 | Unenroll | Explicitly clears the vault key and label. |

## APPENDIX B. REPRODUCIBILITY NOTES

The implementation discussed here is organized as: pico-fido (firmware and emulator), pico-vault-enroller (standalone provisioning helper), pico-keys-sdk (SDK dependency), and picokeyapp (GUI credential provider) [4]. The enroller is intentionally treated as a separate project because its threat model differs from the GUI's threat model. A publication artifact should include commit identifiers, compiler and dependency versions, test logs, certificate-generation fixtures that do not contain production CA secrets, and a clear statement of which live tests were executed on enrolled hardware versus skipped because a required precondition was unavailable.

For contributors, the practical review order is: read the threat model; inspect pico-vault-enroller/vault_enroller.py because it handles plaintext Kvault; inspect src/fido/vault.c and vault.h for parsing, derivation, and zeroization; inspect gui/fido/vault.py and vault_panel.py for policy and opaque persistence; then run tests/pico-fido/test_080_vault.py and picokeyapp/tests/test_vault.py. Pytest markers should distinguish deterministic tests from tests requiring a live enrolled authenticator so that a skip cannot be mistaken for protocol success.

## APPENDIX C. GLOSSARY

**AEAD:** Authenticated encryption with associated data; encryption that also detects unauthorized modification of ciphertext and selected visible header fields.

**Board serial:** The authenticator's device identifier. In the optional identity layer it must exactly match a certificate SAN value.

**Enroller:** The separate, security-critical provisioning program that creates or recovers Kvault and enrolls a board.

**Identity CA:** An optional certificate authority that attests who provisioned which board. It is not the vault, not a synchronization backend, and not part of routine export/import.

**Kvault:** A uniformly random 256-bit root secret for one credential-mobility domain. Its enroller recovery copy is protected by a passphrase-derived Argon2id key.

**PKV1:** The proposal's versioned, authenticated envelope carrying one encrypted credential record and visible authenticated routing metadata.

**Vault ID:** A hash-derived identifier for Kvault. It allows matching without revealing the vault root itself.